\documentclass[10pt,twocolumn]{article}
\usepackage{graphicx}
\usepackage{amsfonts}
\usepackage{amssymb}
\usepackage{float}
\usepackage{fancyvrb}
\usepackage{caption}
\usepackage{pdfpages}
\usepackage{multicol}
\usepackage{multirow}
\usepackage{amsmath}
\usepackage{etoolbox}
\usepackage[margin=0.8in]{geometry}
\usepackage{cite}
\usepackage{lipsum}
\usepackage{mathtools}
\usepackage{cuted}
\usepackage{breqn}

\patchcmd{\thebibliography}{\chapter*}{\section*}{}{}
\begin{document}

\twocolumn[
\begin{@twocolumnfalse}

\begin{center}\textbf{{\Large Bounds on Graviton mass using  weak lensing  and SZ effect in Galaxy Clusters }} \\
{ Akshay Rana$^{a}$ \footnote{montirana1992@gmail.com}, Deepak Jain$^b$, Shobhit Mahajan$^a$, Amitabha Mukherjee$^a$\\
$^a$ Department of Physics and Astrophysics, University of Delhi, Delhi 110007, India\\
$^b$Deen Dayal Upadhyaya College, University of Delhi, Sector-3, Dwarka, New Delhi 110078, India\\
$^1$aksrana92@gmail.com}

\end{center}

\begin{center}
\section*{Abstract}
\end{center}

In General Relativity (GR), the graviton is massless. However, a common feature in several theoretical alternatives of GR is a  non-zero mass for the graviton. These theories can be described as massive gravity  theories. Despite many theoretical complexities in these theories, on phenomenological grounds the implications of massive gravity have been widely used to put bounds on graviton mass.  One of the generic implications of giving a mass to the graviton is  that the gravitational potential will follow a Yukawa-like fall off. We use this feature of massive gravity theories  to  probe the mass of graviton by using the largest gravitationally bound objects, namely galaxy clusters. In this work, we use the mass estimates of galaxy clusters measured at various cosmologically defined  radial distances  measured via  weak lensing (WL)  and Sunyaev-Zel'dovich (SZ) effect. We also uses the model independent values of Hubble parameter $H(z)$ smoothed by  a non-parametric method, Gaussian process. Within $1\sigma$ confidence region, we obtain the mass of graviton $m_g < 5.9 \times 10^{-30}$ eV  with the corresponding Compton length scale  $\lambda_g > 6.82$ Mpc from weak lensing and $m_g < 8.31 \times 10^{-30}$ eV   with  $\lambda_g > 5.012$ Mpc from SZ effect. This analysis improves the upper bound on graviton mass obtained earlier from galaxy clusters. \\

\end{@twocolumnfalse}
\vspace{0.0cm}
]

\section{Introduction}

General theory of Relativity (GR) is one of the most elegant theories  of gravity. It had been introduced only on the basis of theoretical principles before being tested and confirmed by  observations \cite{einstein}. Till date  all of its  predictions have been tested and verified in different limits \cite{grproof}. In the weak field approximation, observations like the precise measurement of the perihelion advance of Mercury \cite{mercury}, deflection of light by the  Sun \cite{sun}, gravitational time delay \cite{str}, equivalence principle \cite{ep}, the Nordtvedt effect in lunar motion \cite{nordtvedt}, frame-dragging \cite{frame} etc., show an impeccable agreement with the observations at  solar system  length scales \cite{solar1}. The recent detection of gravitational waves \cite{gw} as well as time lag measurements on  binary radio pulsar \cite{bp} verified the consistency of  GR even in the strong field limit \cite{strongfield}. However, there is still a  lack of direct observations at large length scales ( $\sim$ Mpc) that can establish the consistency of GR \cite{large}.   Further, there is a need to introduce a dark component in the energy budget of the Universe to make it compatible with cosmological observations \cite{darksector}. This dark sector of the  Universe remains unobserved which in turn could be taken as an opportunity to look for alternatives to  GR at large (cosmological) length scales. The study of any deviation and modification of GR remains an exciting  topic of research and has immense theoretical importance. Many alternative theories like $f(R)$ gravity \cite{fr}, Chameleon theory \cite{chemeleon}, Galileon models \cite{galilion} etc., have been proposed.  For a detailed review of alternative theories of gravity, see ref \cite{alternative,ref1,ref2}.\\

Massive gravity theories are a class of alternative theories  to GR that   can explain the cosmic acceleration without invoking a  dark component in the Universe \cite{mg,lavina}. In 1939, Fierz and Pauli (F\&P) proposed a very elegant theory of massive spin $2$ gravitons, in which they added a mass term to the Einstein-Hilbert action in such a way that GR is recovered when the mass term tends to zero \cite{fandp}. However, this theory of massive gravity failed to reproduce the  results of GR at small scales (specially at solar system  scales). This incompatibility is known as van Dam, Veltman, and Zakharov (vDVZ) discontinuity (1970) in literature \cite{vdvz}. But Vainshtein  (1972) presented  a mechanism where he showed that the vDVZ discontinuity can be cured by  taking into account  non-linearities \cite{vain}.  Soon thereafter, Boulware and Deser (1972)  found that a ghost-like negative kinetic term appears in the non-linear massive gravity theory which destabilizes the background. This instability is known as the BD instability \cite{bd}.  However, in the last decade  de Rham, Gabadadze, Tolley (dRGT 2011)  provided a nonlinear completion to the Fierz-Pauli massive gravity theory that evades the BD ghost instability \cite{drgt}. Recently, LIGO has reported a gravitational wave event GW170718  which rules out many scalar-tensor theories though   many massive gravity models survive this test\cite{gwref1,gwref2}. Presently, the DGP model \cite{dgp}as well as  Bigravity models \cite{bigravity} which support modification of GR at large scales have emerged to address  some fundamental issues in cosmology such as Dark matter, Dark energy, inflation etc. \\

Due to many conceptual and theoretical difficulties, massive gravity theories are yet to emerge as strong contenders for replacing GR. However, various generic phenomenological features of the motivation of these theories can be used to probe graviton mass. Three traditionally used motivations for this purpose are as follows: [for details of all methods see ref.\cite{gravitonmass,graviton2,will97}].\\

a)  In the case of a massive graviton, the left hand side of the basic Poisson equation (i.e. $\nabla^2 \phi$) which governs the gravitational potential in a linear regime, gets modified to  $(\nabla^2-m^2)\phi$ and the corresponding form of gravitational potential gets modified from the Newtonian potential to a Yukawa-line potential. Giving a small mass to the graviton,  at small length scales (solar system scales) the departure of the Yukawa potential from the Newtonian potential would  be very small but   at large distances (galactic and extragalactic scales) it would become significant. \\

b)   A massive graviton would not travel at the  speed of light. This would modify the corresponding dispersion relation. One way to test this is by comparing  the arrival times of a gravitational wave and the  electromagnetic counterpart \cite{gw150916,pulsartiming}.  The most recent and reliable bound on the graviton mass from this approach  is obtained from GW170104 which is  $m_g< 7.7 \times 10^{-23}$ eV \cite{gw170104}.\\

c) The above mentioned  approaches are  straightforward and can  easily be  inferred from the linear theory. However, many massive gravity theory also give rise to a fifth-force kind of interaction due to non-linear effects. Several bounds on graviton mass from  models in this catagory have been obtained by studying the sensitivity of the fifth force effect on  the precession of the  Earth-Moon by using  Lunar Laser Ranging ($m_g<10^{-32}$eV) \cite{llre}, radiated power from binary pulsar systems ($m_g<10^{-27}$eV) \cite{binary} and  structure formation ($m_g<10^{-32}$eV)\cite{park}. However, all these bounds  are restricted to the DGP and dRGT model within their decoupling limit approximations and sensitive to the details of the model. \cite{gravitonmass}. \\

In this work, we use the  fact that in the case of a massive graviton the  gravitational potential due to a static massive object of mass $M$ changes from the  Newtonian to the Yukawa type fall-off and can be  parametrized as
\begin{equation}
V= \frac{GM}{r} e^{-r/{\lambda_g}}
\end{equation} 

where $\lambda_g$ is a length scale that represents the range of  interaction due to exchange of gravitons of mass $m_g= \frac{h}{\lambda_g c}$. 

Hare [1973] first proposed this phenomenological approach of the  probing graviton mass\cite{hare}.  Goldhaber and Nieto (hearafter GN74) used  Galaxy Clusters for the first time to limit the graviton mass. Using  the Holmberg galaxy cluster they   found  the bound on the graviton mass of the order of $m_g < 1.1 \times 10^{-29} $ eV or $\lambda_g > 10^{20} $ km (3.24 Mpc) \cite{goldhaber,holmberg}. Choudhury et. al. (2004) \cite{sm} derived the convergence power spectra of weak lensing under Newtonian as well as Yukawa gravity. To obtain the bound on the graviton mass they compared the corresponding cosmic shear with observations of weak lensing from a cluster. Within a $1\sigma$ confidence region they constrained the graviton mass to 
 $m_g < 6 \times 10^{-32}$ eV or $\lambda_g > 3 \times 10^{21}$ km ($\sim$ 97 Mpc). \\

 Recently,  S. Desai (2017) used the dynamical features of  the Abell 1689 galaxy cluster to probe the graviton mass  \cite{desai} and obtained the  upper limit on the graviton mass of  $m_g< 1.37 \times 10^{-29}$ eV or $\lambda_g > 9.1  \times 10^{19} $ km (2.95 Mpc) within 90\% confidence level.  Zakharov et. al. [2016]  also use a similar phenomenological consequence of massive gravity and show that an analysis of bright star trajectories near the Galactic Center could bound the graviton mass. They found the upper bound for graviton mass from their work to be  $m_g < 2.9 \times 10^{-21}$eV within 90\% confidence level\cite{zakh,zakha}.\\
 
 In this era of precision cosmology with access to detailed observations and improved knowledge, it is desirable to check and explore these limits with extended datasets.  Here we further extend this approach to limit the graviton  mass by analysing the acceleration profile of Newtonian and Yukawa gravity. For this  we use the full catalog of galaxy clusters, obtained by using weak lensing and SZ effect. In the past,  a single galaxy cluster  has been used to constrain the  graviton mass. However, we have not  come across any work in literature where the presently available full catalog of galaxy clusters has been used for this purpose. The structure of the paper is as follows: In Section 2 , we discuss  the datasets and the method of mass estimation of galaxy cluster using weak lensing and SZ effect. The method and the result are discussed in the Section 3 and 4, respectively. Conclusions  are summarized   in Section 5.

\begin{figure*}[]
\centering
\includegraphics[height=6cm,width=8cm,scale=4]{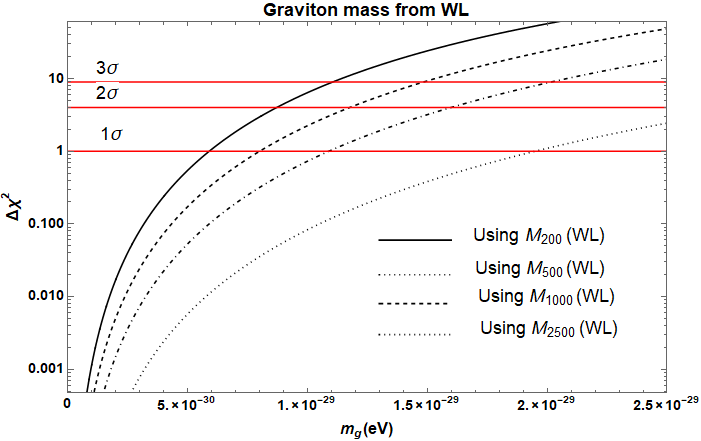}
\includegraphics[height=6cm,width=8cm,scale=4]{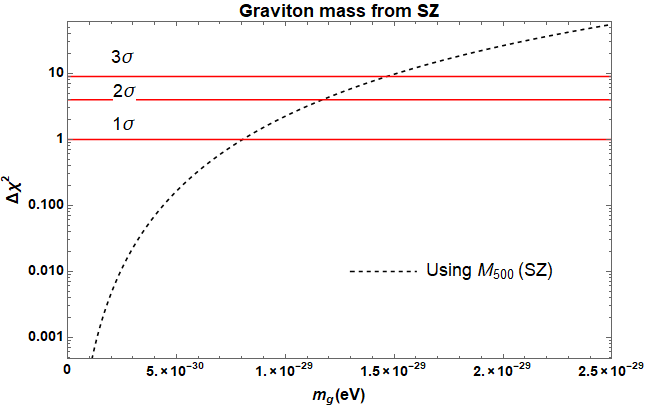}
\caption{ \label{ER1}\footnotesize    $\Delta\chi^2$ is plotted as a function of $m_g$  (in eV) obtained by using the mass of galaxy clusters with weak lensing and SZ effect. The horizontal  red lines at $\Delta \chi^2 = 1, 4$ and $9 $ represent the $1\sigma$, $2\sigma$ and $3\sigma$ confidence limits. \textbf{Left panel} curves have been plotted by using the $M_{200}^{WL}$, $M_{500}^{WL}$, $M_{1000}^{WL}$ and $M_{2500}^{WL}$  estimates of 50 galaxy clusters studied by using weak lensing. In the \textbf{Right panel}, the black dashed line is plotted by using the $M_{500}^{SZ}$ mass estimate of 182 galaxy clusters studied by using SZ effect. (For interpretation
of the colours in the figure(s), the reader is referred to the web version of this article.) }
\end{figure*}

\section{Dataset}

Galaxy clusters are the most massive  gravitationally bound structures that emerge in the large scale structure (LSS) web. Several characteristic features of galaxy clusters like the emission of X-Ray radiation from inter-cluster medium through thermal bremsstrahlung phenomenon and the thermal shift in the black-body spectrum of CMB photons through inverse Compton scattering (SZ effect)  play a crucial role in probing the Universe \cite{cl06}. Weak gravitational lensing of background objects by clusters also provides crucial information about the evolution of large scale structure and composition of the universe \cite{wk99}. We use two different mass measurements of galaxy clusters from the  weak lensing and SZ effect to probe the graviton mass.

\subsection{Mass estimation of Galaxy cluster using Weak lensing}

Weak lensing (WL) is the cleanest method to estimate the mass of  large scale structures because gravitational lensing is  sensitive to the total matter distribution and is not affected by the  physical and dynamical state of clusters. The observable quantity  measured in weak lensing surveys is the  small change in the ellipticity or the tidal distortion of a galaxy's image known as shear. If this shear is caused by large scale structures like clusters, then it is known as cosmic shear. It is directly related to the projected foreground mass of lensing objects enclosed within the cluster radius \cite{bhjain}.

In this work, we use the weak lensing mass measurement of 50 most massive galaxy clusters  (redshift range $0.15<z<0.3$) analysed by Okaba et. al (2015)  in the Local Cluster Substructure Survey (LoCuSS)\cite{okaba,okaba13,okaba14}. To compute the mass of a galaxy cluster,  a model of the shear profile of individual cluster has been fitted to the observational data     by adopting the universally  accepted Navarro, Frank \& White (NFW) mass density profile \cite{nfwprofile} of dark matter halo. The effect of systematic bias in measurement, contamination of background galaxies and intrinsic asphericity of galaxy clusters have also been taken care of in the dataset.


  In this analysis, we use mass estimates of galaxy clusters calculated by using the same approach at radius $R_{200}$, $ R_{500}$, $R_{1000}$ \& $R_{2500}$  and defined as $M_{200}^{WL}$,$ M_{500}^{WL}$, $M_{1000}^{WL}$ \& $M_{2500}^{WL}$ (For data see Table B1 in ref.\cite{okaba}) \footnote{  The quantities  $R_{\Delta}$  $M_{\Delta}$ or else, having subscript $\Delta=$ 200, 500, 1000 and 2500  gives the measure of these quantities at a radius at which the mean density $\rho_{\Delta}$ of cluster is $\Delta$ times the critical mass density $\rho_{c}$ of  the universe at corresponding redshift $z$ of cluster.}

 \subsection{Mass estimation of Galaxy cluster from SZ effect}

The SZ effect accounts for the distortion in the black body spectrum of the  Cosmic Microwave Background (CMB) radiation generated via  inverse Compton scattering of CMB photons by the hot and energetic free electrons in the inter-cluster medium (ICM). The magnitude of  SZ distortion in the  CMB spectrum is measured by a parameter called the  Compton parameter $y$, which is a measure of gas pressure integrated along the line of sight.  The gas pressure is directly related to the gravitational potential of clusters. Hence, the mass of a cluster can be calculated from the SZ signal through the pressure profile of the galaxy cluster \cite{szeffect}. Arnaud et. al. (2010)  proposed a cluster electron pressure profile as a function of radius $r$ of cluster, modeled by using a generalized NFW density profile and named it as the  Universal Pressure Profile (UPP) \cite{arnaud}.
 Hasselfield et. al.(2017)  use this pressure profile to  develop a scaling relation between  the SZ observable and cluster mass and  hence provide an estimate of  the mass of galaxy cluster observed in the Atacama Cosmology Telescope (ACT) \cite{hilton13}.

Recently, Hilton et. al. (2017)  present a catalog of $182$ optically confirmed galaxy clusters   detected via the SZ effect at the Atacama Cosmology Telescope (ACT) survey  in the redshift range $0.1<z<1.4$ \cite{hilton}.  The  clusters' mass $M_{500}$ has been calculated by using the SZ signal under the assumption of the  Universal Pressure Profile of  cluster electron pressure within a radial distance $R_{500}$. In this work, we used these mass estimates of $182$ galaxy clusters and represented them as ($M^{SZ}_{500}$). For details of method see ref. \cite{hilton13, hilton} and for the dataset see Table A.3 in \cite{hilton} with index $M^{UPP}_{500}$.

  \begin{figure*}[]
\centering
\includegraphics[height=6cm,width=8cm,scale=4]{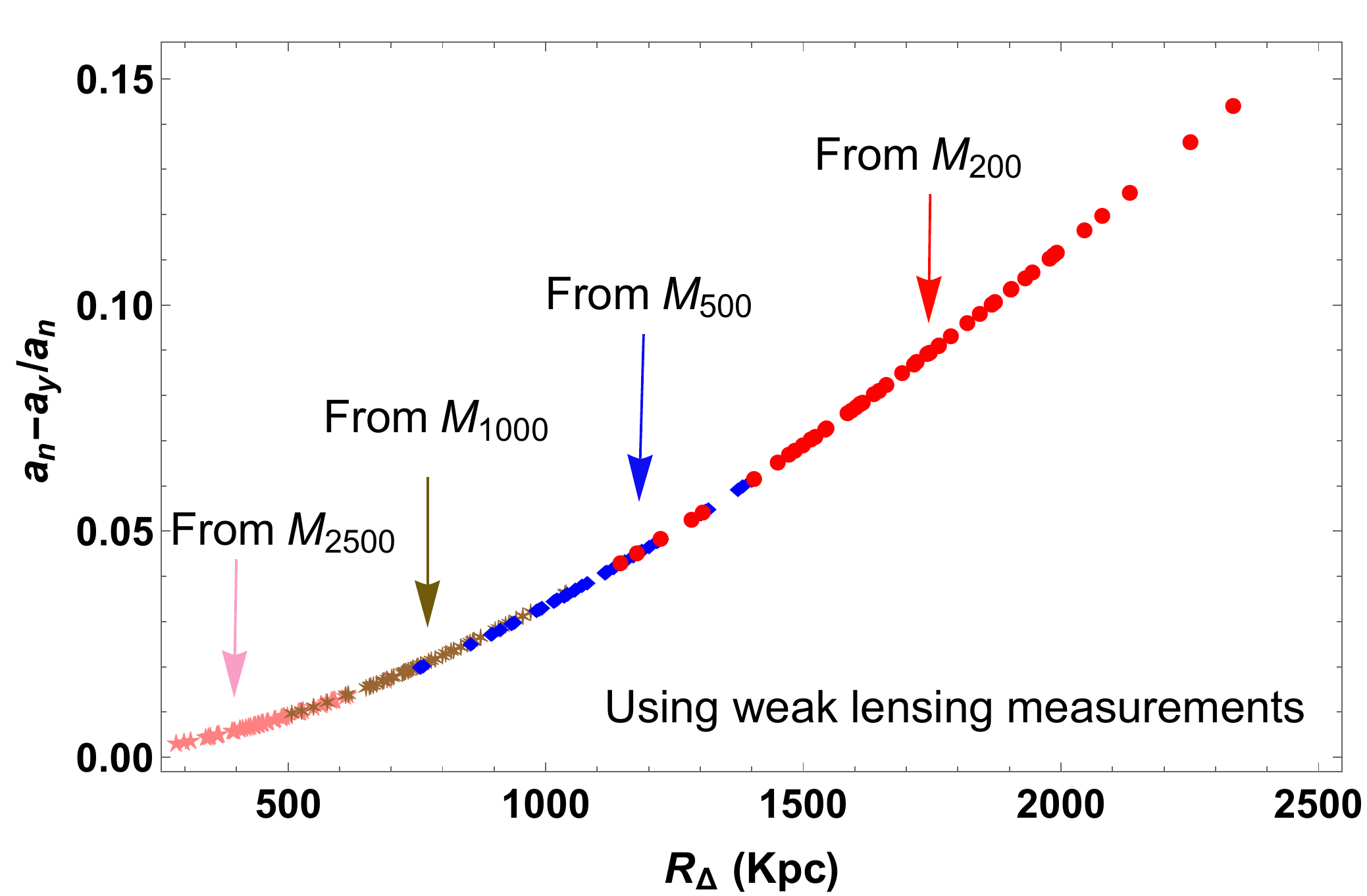}
\includegraphics[height=6cm,width=8cm,scale=4]{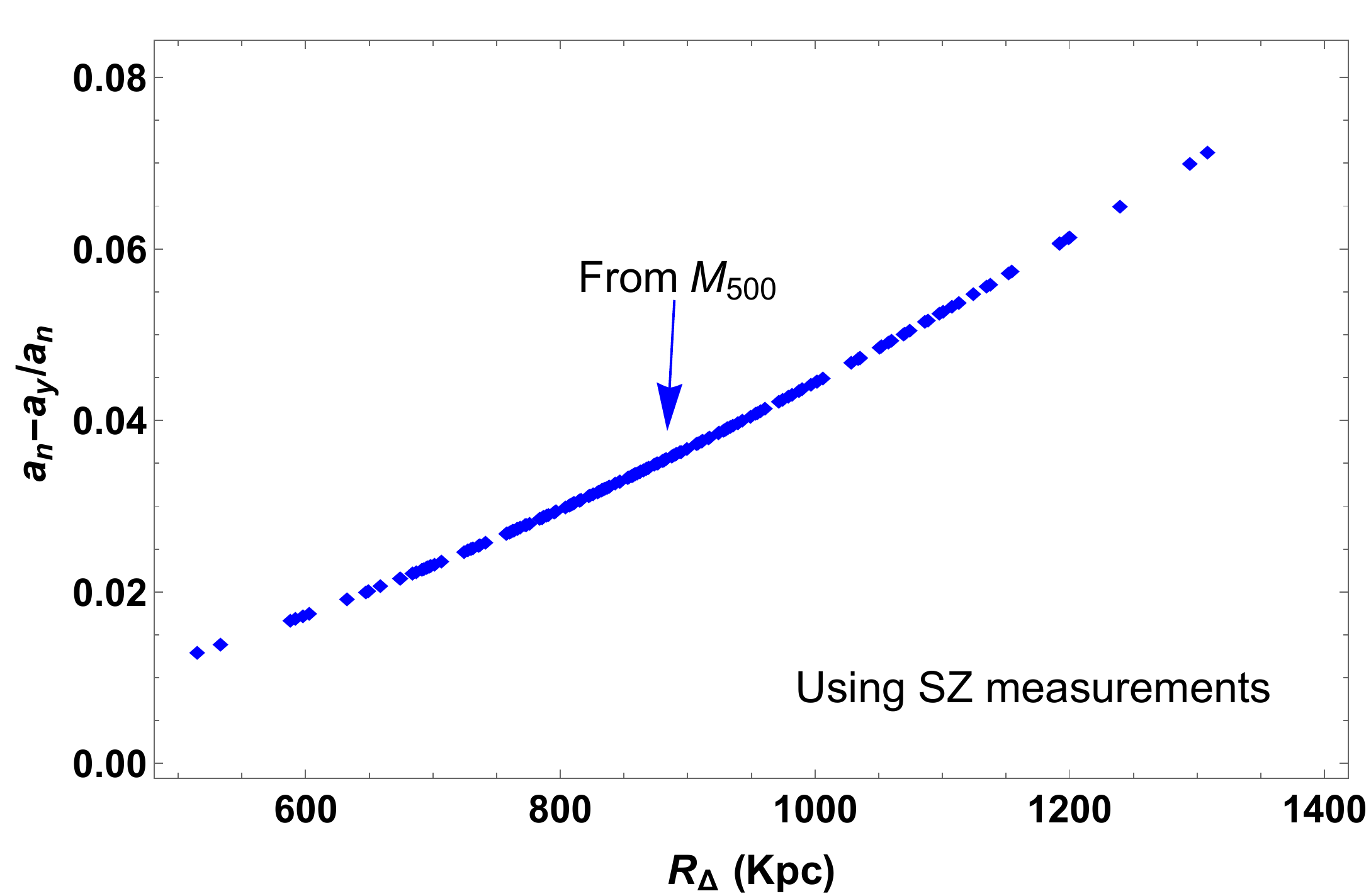}
\caption{ \label{hubbledata}\footnotesize Plot of fractional difference between Newtonian \& Yukawa acceleration profile i.e. $(a_n-a_y)/a_n$ with radial distance $R_{\Delta}$ in kpc. In the \textbf{left panel,}  $(a_n-a_y)/a_n$ vs $R_{\Delta}$ for 50 galaxy clusters studied through weak lensing is plotted. Points represented by the pink star, brown asterisks, blue diamonds \& red dots represent the  $M_{2500}^{WL}$, $M_{1000}^{WL}$, $M_{500}^{WL}$ and $M_{200}^{WL}$ mass measurements of 50 galaxy clusters respectively. In the \textbf{right panel},  $(a_n-a_y)/a_n$ vs $R_{\Delta}$ for 182 galaxy clusters studied through SZ effect is plotted.}
\label{wkdist}
\end{figure*}

\section{Method}

Given the mass of a galaxy cluster at any particular radial distance, the gravitational acceleration, ${a_{n}}$, in Newtonian gravity can be  written as

\begin{equation}
a_{n} = \frac{G M_{\Delta}}{R_{\Delta}^2}
\label{nw}
\end{equation}

where $ M_{\Delta}$  represents the mass of the galaxy cluster within a radius $R_{\Delta}$, which is defined as a distance from the core of cluster at which the density of galaxy cluster becomes  $\Delta$ times  the critical density $\rho_c$ of the Universe at that epoch.  The mass of the galaxy cluster can be defined as;

\begin{equation}
M_{\Delta}=  \Delta \times \rho_c  \times \frac{4\pi}{3} R_{\Delta}^3
\label{mdel}
\end{equation}

The critical density of the Universe is given by $\rho_c = \dfrac{3H^2(z)}{8\pi G}$, where $H(z)$ represents the Hubble parameter.
On the other hand, if we assume a modified theory with massive gravitons,  the corresponding gravitational acceleration at any particular radial distance would take the Yukawa form and we get,

\begin{equation}
a_{y}=   \frac{GM_{\Delta}}{R_{\Delta}}  \exp(-R_{\Delta}/ \lambda_g) \left( \frac{1}{R_{\Delta}} + \frac{1}{\lambda_g} \right)
\label{yu}
\end{equation}

By using Eq. \ref{mdel}, one can rewrite the above mentioned acceleration expressions $a_{n}$ and $a_{y}$ as,

\begin{equation}
a_{n}(z,M_{\Delta})= (G M_{\Delta})^{1/3} \left(\frac{ H^2(z) \Delta}{2}\right)^{2/3}
\label{newt}
\end{equation}

and

\begin{dmath}
a_{y}(z,M_{\Delta},\lambda_g){= (G M_{\Delta})^{2/3}} \left(\frac{ H^2(z) \Delta}{2}\right)^{1/3} \times \\ \exp  \left[ - \frac{1}{\lambda_g}\left(\frac{2M_{\Delta}G}{H^2(z) \Delta}\right)^{1/3}\right] \left[ \frac{1}{\lambda_g}+ \left( \frac{H^2(z) \Delta}{2 M_{\Delta}G}\right)^{1/3}\right]
\label{yuk}
\end{dmath}

In order to put any limit on graviton mass by using these acceleration profiles, we require independent information about $M_{\Delta}$ and Hubble parameter $H(z)$. For $M_{\Delta}$, we use the masses  of galaxy clusters estimated by using the SZ effect and weak lensing properties, as mentioned in the previous section.

\begin{figure}[t]
\includegraphics[height=6cm,width=8cm,scale=4]{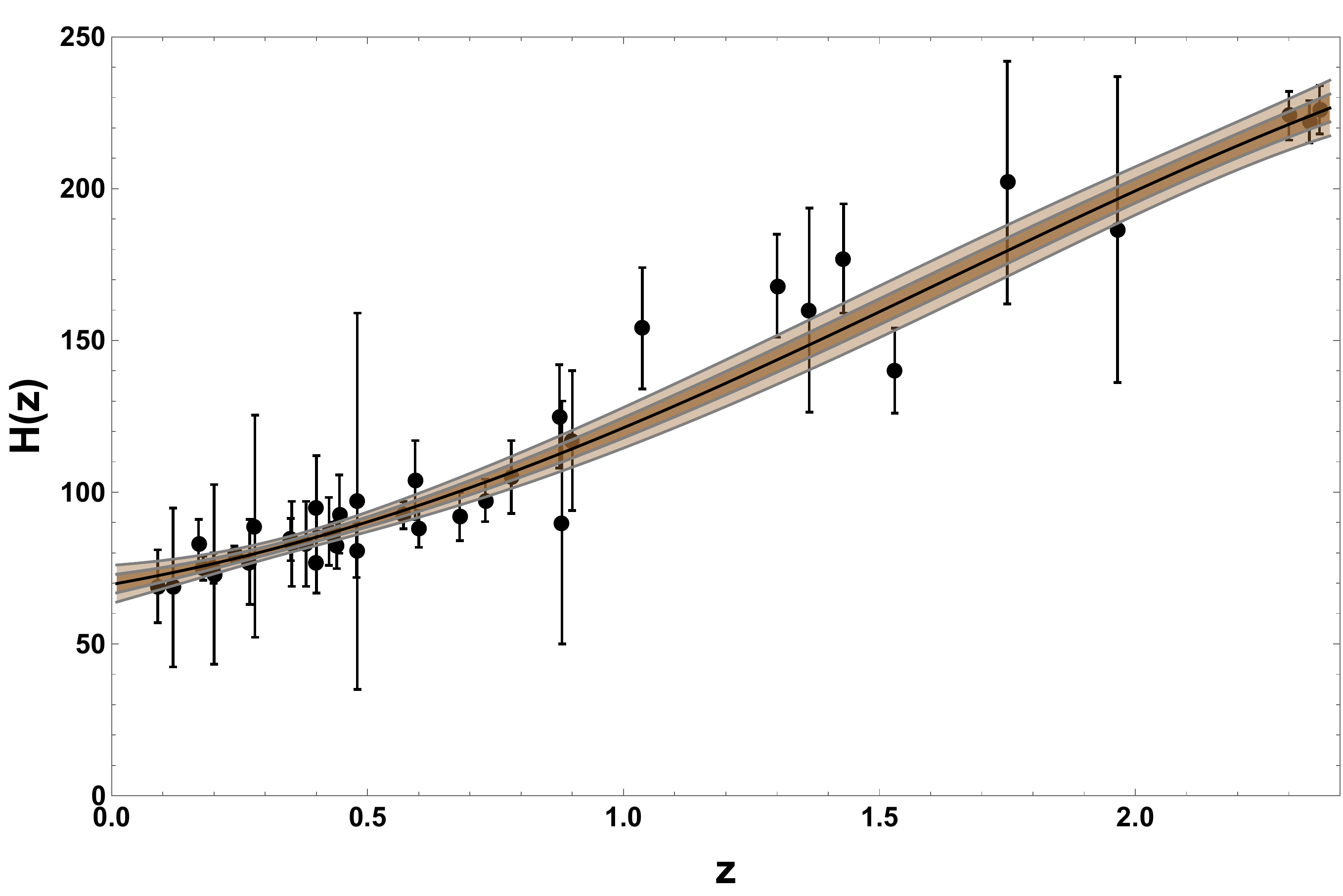}
\caption{{\label{aaa}}\footnotesize  Plot of Hubble Parameter $H(z)$ vs redshift $z$. The black points with error bars are the  $38$  measurements of $H(z)$ and the solid line with $1\sigma$ and $2\sigma$ confidence limits are the corresponding smoothed values of the  $H(z)$ plot by using a model independent non-parametric method, "Gaussian Process".   }

\end{figure}

 To estimate the value of the Hubble parameter $H(z)$  at the corresponding redshift, we use  the $38$ observed Hubble parameter values of  $H(z)$ in the redshift range $0.07<z<2.34$  calculated by using the differential ages of galaxies, Baryonic Acoustic Oscillation (BAO) and Lyman $\alpha$ measurement \cite{hubbledata}. As seen in Fig. \ref{aaa}, We apply a non-parametric technique (Gaussian process) to smoothen it which enables us to find  the  model independent value of $H(z)$ at all desired redshifts of the galaxy clusters \cite{gp0}. [For more details of Gaussian process see \cite{gp0,gp,gp1}].

\begin{table*}[]
\begin{center}
\small{
\begin{tabular}{|l|l|l|l|l|l|}
\hline
\multicolumn{6}{|c|}{\footnotesize Upper Bound on Graviton mass $m_g$ (in eV) and lower bound on  $\lambda_g$ (in Mpc)} \\
\hline
\small Data  & Parameter& 1 $\sigma$ (68.3\%) & 1.64 $\sigma$ (90\%)  & 2 $\sigma$ (95.5\%) & 3 $\sigma$ (99.7 \%)   \\ \hline

\multirow{2}{*}{$M_{200}^{WL}$  }

& $m_g<$ (in eV) & 5.902 $ \times 10^{-30}$ & 7.849  $\times 10^{-30}$ & 8.715  $\times 10^{-30}$ & 1.105 $\times 10^{-29}$  \\
& $\lambda_g>$(Mpc) & 6.822 & 5.132  & 4.622  & 3.643  \\
 \hline

\multirow{2}{*}{$M_{500}^{WL}$ } & $m_g<$ (in eV) & 8.003 $ \times 10^{-30}$  &1.053 $ \times 10^{-29}$ & 1.175 $ \times 10^{-29}$& $1.48 \times 10^{-29}$ \\
 & $\lambda_g>$(in Mpc)& 5.033 & 3.824 & 3.427 & 2.713     \\ \hline

\multirow{2}{*}{$M_{1000}^{WL}$ }& $m_g<$ (in eV) & 1.088 $ \times 10^{-29}$ & 1.427 $ \times 10^{-29}$ & 1.598 $ \times 10^{-29}$ & 2.017 $\times 10^{-29}$ \\
 & $\lambda_g>$(in Mpc) & 3.700 & 2.821  & 2.520  & 1.997     \\ \hline

\multirow{2}{*}{$M_{2500}^{WL}$ } & $m_g<$ (in eV)  & 1.952 $ \times 10^{-29}$ & 2.583 $ \times 10^{-29}$ & 2.894 $ \times 10^{-29}$ & 3.641 $ \times 10^{-29}$ \\
 & $\lambda_g>$(in 	Mpc)& 2.060 & 1.560 & 1.390 & 1.100   \\ \hline
 \hline

 \multirow{2}{*}{$M_{500}^{SZ}$  }

& $m_g<$ (in eV) & 8.307 $ \times 10^{-30}$ & 1.051  $\times 10^{-29}$ & 1.169  $\times 10^{-29}$ & 1.461 $\times 10^{-29}$  \\
& $\lambda_g>$(Mpc) & 5.012 & 3.831 & 3.443  & 2.747  \\
 \hline

\end{tabular}}
\end{center}
\caption{\footnotesize  Bounds on the graviton mass $m_g$ and corresponding Compton length scale $\lambda_g$  within   $1\sigma$, $1.64\sigma$, $2\sigma$ and $3\sigma$ confidence limits estimated by using  $M_{200}^{WL}$, $M_{500}^{WL}$, $M_{1000}^{WL}$, $M_{2500}^{WL}$ and $M_{500}^{SZ}$   }
\label{gravitonbound}
\end{table*}

Once the acceleration corresponding to the Newtonian potential and Yukawa potential are known, we define a chi-square $\chi^2$ as;

\begin{equation}
\chi^2 = \sum_{i} \left[\frac{a_{n,i}(z,M_{\Delta})-a_{y,i}(z,M_{\Delta},\lambda_g)}{\sigma_{a,i}}\right ]^2
\label{chisquare}
\end{equation}

where $\sigma_a$ gives the error in acceleration obtained by adding the errors of mass estimate, $\sigma_{M_{\Delta}}$ and Hubble parameter $\sigma_H$ in quadrature, given by,

\begin{equation}
\sigma_a = \frac{a_n}{3}\sqrt{\left( \frac{\sigma_{M_{\Delta}}}{M_{\Delta}} \right)^2 + 16\left(  \frac{\sigma_{H}}{H(z)} \right)^2 }
\end{equation}

 In this analysis, we have only one model parameter, $\lambda_g$, related to the graviton mass by the expression, $\lambda_g= h/m_gc$ and the $\chi^2$ has been summed over all the datapoints present in catalog. It can  be seen easily that as $\lambda_g \rightarrow \infty$ or  $m_g \rightarrow 0$ then  $a_{y}(z,M_{\Delta},\lambda_g)$ reduces to $a_{n}(z,M_{\Delta})$ and $\chi^2 \rightarrow 0$.  Hence it is obvious that the best fit value of $m_g$ for which $\chi^2$ would minimize (i.e. $\chi^2_{min}$) is zero. To get a bound on  graviton mass with  $68.3\% $ or $1 \sigma$ confidence we put a threshold limit $\Delta \chi^2 < 1.0$, where $\Delta \chi^2= \chi^2- \chi^2_{min}$. Similarly for $90.0 \% $  ($1.64\sigma$), $95.5\% $ ($2\sigma$) and $99.7\% $ ($3\sigma$) confidence limits we have threshold limits at $\Delta \chi^2= 2.71, 4$ and $9$ respectively (see Fig. \ref{ER1}).
 
\section{Result}

 The deviation of the Newtonian potential at large scales depends  on how the total mass of galaxy cluster changes with respect to the radial distance. Presently, the  weak lensing and SZ effect  related surveys enable us to find out the mass of these large celestial objects, i.e. galaxy clusters with great precision at a radial distance where all the clusters have the same density. In this work, we use  two such catalogs, in which first one contains the  mass estimates of 50 galaxy clusters calculated by using the weak lensing properties and defined as  $M_{200}^{WL}$, $M_{500}^{WL}$, $M_{1000}^{WL}$ and $M_{2000}^{WL}$. The  second catalog contains the  mass estimate of 182 galaxy cluster derived by using the SZ property of clusters and defined as $M_{500}^{SZ}$ [Note:  ``Universal Pressure Profile" (UPP) of pressure distribution within the cluster has been used to estimate mass ]. In the notation of mass, the subscript contains a number ($\Delta= 200, 500, 1000 $ etc.), which represents the radial extent of cluster upto a region where the density of cluster is $\Delta$ times of the critical density of the universe.

   The mass estimates of galaxy clusters  indirectly depend upon the form of the potential.  It requires  input about the mass profiles for dark matter halos. In both datasets used in this analysis,  dark mass distribution have been assumed to follow the NFW profile. This is an empirical mass profile identified in N-body simulations  of structure formation and widely accepted in the literature \cite{nfwp,nfw2}. Since these simulations implicitly assume a Newtonian potential, it would not be quite correct to claim that this analysis is completely independent of any cosmological assumption.

As deduced from  Fig. \ref{ER1}, Table \ref{gravitonbound} presents the bounds on the graviton mass obtained through different mass estimates of galaxy clusters at different confidence limits ($1\sigma$, $1.64\sigma$, $2\sigma$ \& $3\sigma$). One obvious point that can be directly inferred from  Fig. \ref{wkdist} as well as from Table. \ref{gravitonbound} is that as the length scale increases, the fractional change between the Newtonian \& Yukawa acceleration profile $(a_n-a_y)/a_n$ becomes significant. In the left  panel of Fig. \ref{wkdist},  we explore this fractional change by using WL upto 2.3 Mpc length scale, where it becomes quite significant ( approximately 15\%). Similarly, in the right panel of Fig. \ref{wkdist}, this difference is approximately 7\% at a radial distance  1.3 Mpc which is calculated by using the mass measurements of 182 galaxy clusters studied through the  SZ effect. As expected this behaviour of increasing difference between the two potentials with increasing distance has  been shown by all clusters  irrespective of their redshift.

In this work,  the strongest bound  on the graviton mass  obtained is    $m_g< 5.9 \times 10^{-30}$ eV or $\lambda_g > 6.822$ Mpc  by using $M_{200}^{WL}$ estimate of $50$ galaxy clusters within a $1\sigma$ confidence limit. The catalog having $182$ independent mass measurements $M_{500}^{SZ}$ of galaxy clusters, derived by SZ effects constrains the graviton mass to  $m_g< 8.30   7\times 10^{-30}$ eV or $\lambda_g > 5.012 $ Mpc. We also notice  that the graviton mass is seen to be  sensitive to the length scale. It is clear from the bounds given in Table 1 that the graviton mass bounds  improve as one moves  out in the radial distance (i.e. from $R_{2500}$ to $R_{200}$).

\section{Conclusion}

 Galaxy clusters are the largest gravitationally bound objects  in the Universe and occupy a special place in the hierarchy of cosmic structures. The  observational characteristics of galaxy clusters can be extensively used to study the properties of the cluster galaxy population and those of the hot diffuse intra–cluster medium (ICM). In recent times many dedicated surveys, like Chandra, Newton, ACT, LoCuSS  etc. have studied  the X-ray properties , SZ effect and weak lensing properties of galaxy clusters. There has also been extensive work  to probe different features of cosmology and alternative theories. However, there has not been much work on using galaxy clusters to test  gravity and probe the graviton mass. The very first proposal of using galaxy cluster for probing graviton mass was given by Goldhaber and Nieto (GN74) \cite{goldhaber}. They gave a rough estimate of graviton mass by assuming  that the orbits of galaxies within galaxy clusters are gravitationally bound \& closed and that the distance of outermost galaxies from the core of the Holmberg galaxy cluster is 580 kpc. GN74 further assumed that only Newtonian potential could give rise to such closed and bound orbits and obtained an upper limit on $m_g< 1.1 \times 10^{-29}$ eV. We call it  a rough estimate because neither any statistically significant confidence limit had been defined on the upper limit nor any details about the complicated dynamics of galaxy cluster had been taken into account. The assumption that only a Newtonian potential can give  bound and close orbits has been invalidated in several recent works \cite{boundorbit,nonnewt,bart}.  Desai (2017)  has made an effort to overcome these limitations of GN74 and improve the bounds on graviton mass from a  galaxy cluster \cite{desai}. He used a single galaxy cluster Abell 1689 in order to compare the acceleration profile of the galaxy cluster calculated under both Newtonian as well as Yukawa framework. To find  the mass enclosed within the galaxy cluster at any particular radius, different mass distribution profiles of Dark matter, baryonic matter and galaxy distribution have been used  \cite{abell1689,dmp,galdis,barmass}. However both of these analysis use only a single galaxy cluster which also include many features of the models. We have made an effort to use the complete catalog of galaxy clusters and to obtain  significantly stronger bounds on the mass of graviton.\\
  
  In this work we also extended the same approach further and use the difference between the Newtonian and Yukawa acceleration profile to put bounds on graviton mass by using the mass estimates of galaxy clusters obtained by using weak lensing and SZ effect.  However, it is important to emphasize that this analysis is not completely independent of any cosmological assumption. This is because these mass estimates are obtained by using  the NFW and generalized NFW density profiles of dark matter halos. These density profiles are the outcome of N-body simulations of structure formation which are performed under a Newtonian framework.  The ideal way out to overcome this limitation would be  to run the N-body simulations in Yukawa gravity and refit all scaling relations and obtain separate mass estimates of galaxy clusters in the  Yukawa gravity framework. But the  N-body simulations in Yukawa gravity  need the graviton mass as a prior input or else one would get the  output as a function of the graviton mass. Hence it may not be much of a help in constraining the graviton mass in a self consistent way.  Running such simulations at cluster scales is complicated since   to use the  Yukawa potential, we have to deal with multiple model parameters and density profiles which will further make the process  highly sensitive to multiple parameters. We would like to add that in this work we have followed previous work where alternative gravity models (  modified gravity models \cite{mod1,mod2,mod3}, chameleon gravity \cite{chem1,chem2}, Galileon gravity \cite{gal} etc.) have been tested using galaxy cluster observations with a similar dependence on dark matter halo density profiles obtained from N-body simulations. \\

   \begin{table*}[]
\begin{center}
\footnotesize{
\begin{tabular}{|l|l|l|}
\hline
\multicolumn{3}{|c|}{Various bounds on graviton mass} \\
\hline
\small Hypothesis & \hspace{5cm}\small  Method & \small $m_g$ (in eV)    \\ \hline

\multirow{5}{*}{ Yukawa potential} &  1$\sigma$ bound from weak lensing power spectrum of cluster at z= 1.2 \cite{sm} & $6.00 \times 10^{-32}$  \\

 &  Using  Holmberg galaxy cluster  by assuming scale size around 580 kpc \cite{goldhaber} & $1.10 \times 10^{-29}$ \\
 &  1.64$\sigma$ (90\%) bound  from galaxy cluster Abell 1689 \cite{desai}  & $1.37 \times 10^{-29}$  \\
 &  2$\sigma$ bound  from the precession of Mercury \cite{finn} & $7.20 \times 10^{-23}$  \\

&  1.64$\sigma$ (90\%) bound  using trajectories of  S2 stars near the galactic center \cite{zakh}  & $2.91 \times 10^{-21}$  \\

&  \textbf{1$\sigma$ bound from $M_{200}^{WL}$ mass estimate of 50 galaxy cluster (This work)}& $5.90\times 10^{-30}$  \\

&  \textbf{1$\sigma$ bound from $M_{500}^{SZ}$ mass estimate of 182 galaxy cluster (This work)}& $8.31\times 10^{-30}$  \\
  \hline

\multirow{4}{*}{ Dispersion Relation}

& 90\% upper limit from GW150914 \cite{gw150916} & $1.20 \times 10^{-22}$  \\

& 90\% upper bound from binary pulsar observations \cite{pulsartiming} & $7.60 \times 10^{-20}$  \\

& 90\% upper limit from GW170104 \cite{gw170104} & $7.70 \times 10^{-23}$  \\

& By studying the impacts of a $m_g$ on the B-mode polarization of CMB \cite{lin}& $\sim 9.7 \times 10^{-33}$\\




\hline
\end{tabular}}
 \end{center}
 \caption{\footnotesize Some robust  bounds on graviton mass $m_g$ in eV obtained by using the phenomenological implications of massive gravity theories. For some more interesting works and detailed review see \cite{gravitonmass,llre,zakh,saurya,pastana}  }
 \label{bounds}
 \end{table*}

  The main conclusion of our work is as follows;

  $\bullet$  Instead of using a single galaxy cluster, (as used by Desai (2017) and GN74), we use the presently available observational catalogs  of mass measurements of galaxy clusters  at different cosmologically defined radii.  The use of the complete catalogue including hundreds of independent observations averages out  any possible random error contribution, which is not possible when using a single cluster.

   $\bullet$ We write down the acceleration profile under Newtonian and Yukawa gravity, in terms of the cluster mass and Hubble parameter. The mass measurements of clusters are obtained through weak lensing and SZ effect scaling relations of galaxy clusters. These mass estimates are more reliable because the weak lensing and SZ effect  of galaxy clusters  provide  direct observational results with least input assumptions about cluster dynamics. Moreover, no modelling or assumptions are required about the baryonic mass profile and galaxy distribution within the galaxy cluster in weak lensing studies of clusters, as required for instance  in Desai's work. However, it would not be quite correct to claim that this analysis is completely model independent because the NFW mass profile for dark matter halo has been used  to derive the scaling relations.

   $\bullet$ We also need the model independent values of the Hubble parameter $H(z)$ corresponding to each cluster. To find out  $H(z)$ at the redshift of each cluster, we smoothen  the observational dataset of $H(z)$   by using a model independent non-parametric smoothing technique, \textit{Gaussian process} [see Fig. \ref{aaa} ]. One point to be noted here is that the Hubble parameter enters the analysis through the expression for the  critical density $\rho_{cr}$ of the universe, which is an outcome of a flat FLRW universe. But in order to calculate the value of  $H(z)$, we don't use $\Lambda CDM$ or any other cosmological model.

  $\bullet$  The Yukawa potential decreases rather quickly in comparison to the Newtonian potential. For a given mass of the graviton of the order $m_g \sim 10^{-29}eV$ (i.e. $\lambda_g \sim 4$ Mpc), the fractional difference between the Yukawa and Newtonian acceleration is approximately $ 10^{-14}$ at a length scale of $1$ pc, $\sim 10^{-10}$ at a length scale of $100$ pc, $\sim 10^{-6}$ at a length scale of $10$ kpc and of the order of $10^{-2}$  at a length scale of $1$ Mpc.  Hence, it becomes important to extend this analysis upto very large  (Mpc) length scales to observe a significant difference between these potentials. In this work, the extended catalogs of galaxy clusters gives us an opportunity to probe the graviton mass by using the length scale  beyond 2 Mpc, where the fractional difference between both potentials widens upto 15\% and beyond. This  is a significant improvement as compared to earlier studies.\\

   In Table.\ref{gravitonbound}, we mention all possible bounds obtained in this analysis with different confidence levels (i.e. 68.3\%, 90.0\% 95.5\% and 99.7\% corresponding to $1\sigma$, $1.64\sigma$, $2\sigma$ and $3\sigma$). In Table \ref{bounds}, we summarize the bounds obtained on $m_g$ using phenomenological approach. We finally conclude that;

   $\bullet$ In this work, the strongest bound on graviton mass is obtained by  using the catalog containing 50 independent mass measurement of galaxy cluster $M_{200}^{WL}$ from weak lensing at a radial distance $R_{200}$.  It is $m_g < 5.9\times10^{-30}$eV with the corresponding Compton length $\lambda_g > 6.822$ Mpc within $1\sigma$ confidence interval. The bound on $m_g$ from the catalog having 182 mass measurements of  galaxy clusters $M_{500}^{SZ}$ is $m_g< 8.307 \times 10^{-30}$ eV or $\lambda_g> 5.012$ Mpc.

   $\bullet$ The bound on graviton mass estimated by using the $M_{500}^{WL}$ and $M_{500}^{SZ}$ are almost similar which seems to indicate some degree of  consistency in the mass measurement methods (WL \& SZ) of galaxy clusters.

  With the ongoing and future surveys, our understanding of mass distribution in large scale structures  like galaxies, clusters, super-clusters and filaments will  improve and  more reliable and precise bounds can be obtained using a similar method.

\section*{Acknowledgments}
 A.R. acknowledges support under a CSIR SRF scheme (Govt. of India, FNo. 09/045(1345/2014-EMR-I)). A.R.  thanks Prof. T. R. Seshadri for useful suggestions and  IUCAA Resource Center, University of Delhi for providing research facilities. Authors are also thankful to referee and S. Desai(IIT, Hyderabad) for useful comments.

\end{document}